\documentclass[twocolumn, amssymb, aps, prb, superscriptaddress]{revtex4-2}

\usepackage{graphicx}
\usepackage{dcolumn}
\usepackage{bm}
\usepackage{amsmath}
\usepackage{multirow}
\usepackage{braket}

\usepackage[normalem]{ulem}

\usepackage{url}
\def\lam{\lambda}
\def\tlam{\tilde{\lambda}}
\def\teta{\tilde{\eta}}
\def\k{\kappa}
\def\bk{\bm{k}}
\def\bkp{\bm{k}^{\prime}}
\def\bG{\bm{G}}
\def\br{\bm{r}}
\def\brp{\bm{r}^{\prime}}
\def\rmt{r_{\kappa, \text{MT}}}
\def\ve{\varepsilon}
\def\vef{\varepsilon_{\text{F}}}

\usepackage{xcolor}

\begin{document}
\title{Rigid muffin-tin approximation in
plane-wave
codes for fast modeling of phonon-mediated superconductors}%

\author{Danylo~Radevych}%
\email{dradevyc@gmu.edu}
\affiliation{Department of Physics and Astronomy,
George Mason University, Fairfax, VA 22030, USA}
\affiliation{Department of Physics, Applied Physics and Astronomy,
Binghamton University-SUNY,
Binghamton, New York 13902, USA}

\author{Tatsuya~Shishidou}%
\affiliation{Department of Physics,
University of Wisconsin-Milwaukee,
Milwaukee, WI 53201, USA}

\author{Michael~Weinert}%
\affiliation{Department of Physics,
University of Wisconsin-Milwaukee,
Milwaukee, WI 53201, USA}

\author{Elena~R.~Margine}%
\affiliation{Department of Physics, Applied Physics and Astronomy,
Binghamton University-SUNY,
Binghamton, New York 13902, USA}

\author{Aleksey~N.~Kolmogorov}%
\affiliation{Department of Physics, Applied Physics and Astronomy,
Binghamton University-SUNY,
Binghamton, New York 13902, USA}

\author{Igor~I.~Mazin}%
\email{imazin2@gmu.edu}
\affiliation{Department of Physics and Astronomy,
George Mason University, Fairfax, VA 22030, USA}
\affiliation{Quantum Science and Engineering Center, George Mason University,
Fairfax, VA 22030, USA}

\date{\today}%

\begin{abstract}
We present a pseudopotential-based plane-wave implementation of the rigid muffin-tin approximation (RMTA), offering a computationally efficient alternative to its traditional use in all-electron codes. This approach enables the evaluation of angular-momentum-resolved electron-phonon matrix elements and McMillan-Hopfield parameters of not only elemental transition metals but also their compounds. The results are benchmarked against full-potential linearized augmented plane wave calculations, showing excellent agreement. We further outline a practical route to extract atom- and symmetry-type-resolved electron-phonon coupling constants. By enabling the use of RMTA descriptors within high-throughput workflows, this framework significantly lowers the computational cost of screening candidate superconductors, providing a valuable tool for materials discovery.
\end{abstract}

\maketitle



\section{Introduction}

Historically, the first practical method for calculating the electron-phonon coupling (EPC) constant in metals was rigid muffin-tin approximation (RMTA) \cite{GaspariGyorffy1972, Evans1973, Gomersall1974Aug}, in which the
electronic potential within a sphere around each atom, called the
muffin-tin (MT) sphere,  was assumed to follow ionic
displacements rigidly.
Since the RMTA was
formulated in terms of
MT potentials and
wavefunctions
expanded in radial
functions and spherical harmonics,
it was naturally implemented in many all-electron methods, such
as the augmented plane wave (APW) \cite{Klein1974, Slater1937, Martin2004,
Dewhurst, Gulans2014},
Korringa-Kohn-Rostoker (KKR) \cite{Korringa1947, Kohn1954, Butler1979},
and linear
muffin-tin orbital (LMTO) \cite{Pettifor1977, Andersen1975, Martin2004}
methods,
where
atomic functions inside the MT sphere are evaluated explicitly.
The RMTA method worked
reasonably well for materials with localized $d$-orbitals and close-packed structures (e.g., V, Nb, Mo, and Pd) \cite{Pettifor1977, Papaconstantopoulos1977,
Butler1977, Pickett1982, Ginzburg1982, Mazin1982, Mazin1984,
Bose1990},
and was actively used to study superconductivity of various materials, such as $A15$ \cite{Klein1979, Pickett1982} and $C15$ \cite{Jarlborg1980} compounds, hydrides \cite{Papaconstantopoulos1975}, nitrides \cite{Pickett1981}, etc.
While this method was designed with transition metals in mind, and routinely tended to underestimate the EPC in $sp$-metals, corrections accounting for the long-range tails of the ionic MT potential at large distances in such materials have been proposed  \cite{Zdetsis1981, Mazin1984}
and successfully applied to, e.g., Ag \cite{Mazin1984} and Al, Ga, and Pb \cite{Zdetsis1981}.

Later on, the RMTA was superseded by
the much more accurate and universal
linear-response method, formulated within density-functional perturbation theory (DFPT),
\cite{Giannozzi1991, Gonze1997, Baroni2001, DalCorso2001}
and by
superconducting density-functional theory (SCDFT)
\cite{Oliveira1988, Luders2005, Marques2005, Sanna2020}, both of which are
well-suited
for fast pseudopotential-based plane-wave (PS$+$PW) codes,
and are widely used for predicting superconducting properties.
The first high-throughput materials screenings involving hundreds of $\alpha^2F(\omega)$-based superconductivity calculations have been recently reported~\cite{Choudhary2022, Gibson2025,Nepal2024,Gibson2025b}, but the high DFPT computational cost imposes limitations in the form of restricted unit cell sizes or confined composition spaces. While these efforts mark important milestones in applying machine learning to guide superconductor discovery, larger datasets, producible with fast physics-based methods, are needed for identifying broader design rules.

Several fast descriptors for superconductors have been proposed, typically tailored to particular materials families. For MgB$_2$-type layered compounds, the difference between in-plane boron phonon frequencies at the M and $\Gamma$ points was used as a softening marker expected to correlate with the EPC \cite{Kolmogorov2008}. In $sp$-metals, scaled EPC constants evaluated at the Brillouin-zone center were introduced as a proxy for the total EPC \cite{Rodriguez1990, Sun2022}, but this approach cannot describe cases where zone-boundary phonons play a significant role. More recently, frozen-phonon calculations at a small number of high-symmetry $q$-points combined with energy band shifts were proposed to estimate the EPC across materials of general chemistry, with particular emphasis on phases with large unit cells \cite{Dicks2024}, although to date the method has been applied exclusively to $sp$-bonded metals and hydrides. Real-space descriptors have also been explored, including the electron-localization function \cite{Mauro2024, Belli2025} and spatial derivatives of the Kohn–Sham potential \cite{Novoa2025}, which were used to estimate $T_c$ in hydrogen-based superconductors. While effective within their target classes, usually comprised of light elements, these descriptors underscore the need for approaches that can extend to transition metals and their compounds.

The RMTA could address this problem since it has been
successfully applied
to transition metals, but so far it has not been
formulated in a way
suitable for
pseudo\-potential-based
codes,
which are preferably used by
modern search algorithms.
To assist future superconductor predictions
and enhance the
functionality of such codes, the current work
presents a practical approach for evaluating EPC constants using the RMTA within the plane-wave pseudopotential formalism.
Developed method,
which is a fast postprocessing step on top
of a standard self-consistent field (SCF)
density functional theory (DFT) calculation,
is applied to
transition metals and their compounds.
Calculated
electron-phonon (EP) matrix elements \cite{Pettifor1977}
and McMillan–Hopfield (MH) parameters \cite{McMillan1968, Hopfield1969, Gomersall1974Aug, Pettifor1977, Ginzburg1982}
are shown to agree well with the RMTA implementation
within the
full-potential linearized augmented plane wave (FLAPW)
method.

The paper is organized as follows:
Sec.~\ref{sec:method} introduces the notations and describes the methodology, with more detailed derivations provided in the Appendices.
Sec.~\ref{sec:details} outlines relevant technical aspects of the tests.
Calculated partial EP matrix elements and MH parameters are compared to those obtained with an FLAPW code in Sec.~\ref{sec:results}.
In the same section, estimates of EPC constants for simple metals are obtained using experimental Debye temperatures.
Sec.~\ref{sec:discs} discusses how the RMTA descriptors can be leveraged with machine learning algorithms.
Unless stated otherwise,
atomic Rydberg units are assumed throughout
the paper.

\section{\label{sec:method}Methodology}

This section describes three underlying
assumptions involved in
the RMTA
\cite{Mazin1990, Wiendlocha2006}
along with subtleties of atom-
and type-resolved
EPC constants and MH parameters,
generalizes
standard RMTA equations to
the case of
multiple symmetry types,
introduces partial EP matrix
elements and partial
electronic density of states (eDOS)
needed to calculate MH parameters,
and summarizes key equations
derived in this work
that make the RMTA compatible with
pseudopotentials.

\subsection{Assumption 1: Rigid-ion approximation}
\begin{figure}[!t]
  \includegraphics[width=0.7\columnwidth]{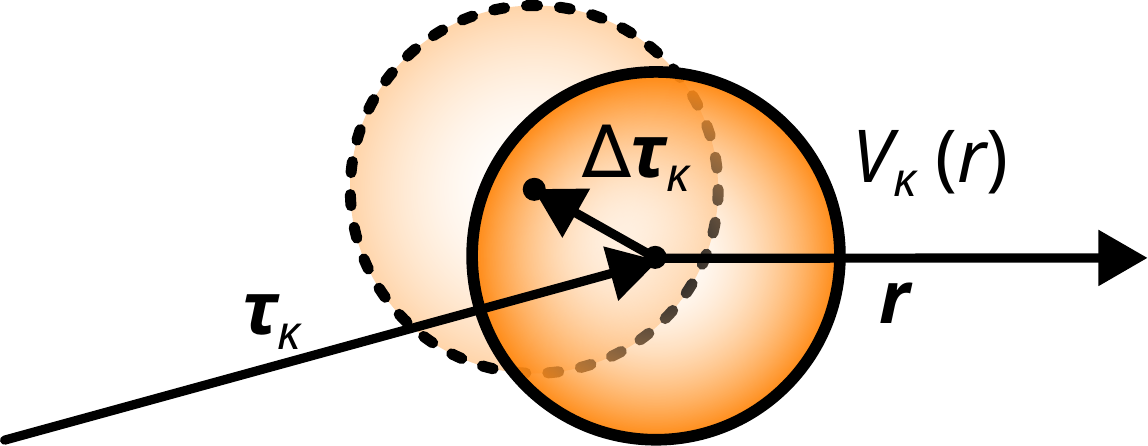}
  \caption{
  Vector definitions
  for Eq.~\eqref{eq:rigid_shift}.
  }
  \label{fig:rmta_def}
\end{figure}

The RMTA relies on the assumption
that
a change in the potential
acting on an electron located at the vector $\br$
from the ionic center $\bm{\tau}_\k$
is given by the rigid displacement
of the electron-ion interaction potential
$V_\k(\br)$
(see Fig.~\ref{fig:rmta_def}),
given by the dot product of the ionic displacement
$\Delta \bm{\tau}_\k$
and gradient of
the potential $\bm{\nabla} V_\k$.
Using the spherical part of the potential, $V_\k(r)$,
present work evaluates this change
in Eq.~\eqref{eq:rigid_shift},
as was done in the original
RMTA formulations \cite{GaspariGyorffy1972, Evans1973, Pettifor1977}.

\begin{equation}
  \begin{aligned}
  \Delta V_\k(r)
  \approx
  - \Delta \bm{\tau}_\k
  \cdot \bm{\nabla} V_\k(r)
  =
  -
  \left(
  \Delta \bm{\tau}_\k \cdot \hat{\br}
  \right)
  \frac{d V_\k(r)}{d r}.
  \end{aligned}
  \label{eq:rigid_shift}
\end{equation}

\subsection{Assumption 2: Local-vibration approximation}

In the RMTA, the EP interaction is treated as local in real space \cite{GaspariGyorffy1972, Gomersall1974Aug},
such that the
total EPC
constant for the whole material, $\lambda$,
can be represented as
a sum of
atomic
EPC constants  $\tlam_\k$ over $N_{\text{atoms}}$
atoms in the considered cell \cite{Papaconstantopoulos2020, Papaconstantopoulos2023}:
\begin{equation}
  \begin{aligned}
  \lambda &=
  \sum_{\k}^{N_{\text{atoms}}}
  \tlam_\k.
  \label{eq:lambda_kappa}
  \end{aligned}
\end{equation}
The EPC constant for each atom is
expressed as a ratio [see Eq.~\eqref{eq:lambda_kappa_def}]
of an
``electronic'' contribution,
represented by
MH parameter $\teta_\k$,
and a ``phonon''
part, given by the
mass of the ion $M_\k$
times its average phonon frequency
$\left< \omega_\k^2  \right>$
\cite{McMillan1968, Hopfield1969, Gomersall1974Aug, Ginzburg1982}.
\begin{equation}
  \tlam_\k =
  \frac{\teta_\k}{M_\k}
  \left<
  \frac{1}{\omega_\k^2}
  \right>
  \approx
  \frac{\teta_\k}{M_\k \left< \omega^2_\k \right>}.
  \label{eq:lambda_kappa_def}
\end{equation}

Although there are known approaches
\cite{Gomersall1974Aug, Foulkes1975, Ginzburg1982} to avoid
full phonon calculations to obtain
$\left< \omega_\k^2  \right>$,
their implementation
is out of the scope of the present work.
Instead, as an example, we approximate $\left< \omega_\k^2 \right>$ in simple metals
using experimental Debye temperatures $\theta_{\text{D}}$,
via the relation $\left< \omega_\k^2 \right> \approx \frac{1}{2} \theta_{\text{D}}^2$.

The main focus of this work is the
derivation of the expressions for the ``electronic'' part, which are more analytically involved
than those for the ``phonon'' part,
as they directly depend on pseudopotentials
and atomic functions derived from them.
The corresponding atomic
MH parameter, $\teta_\k$,
is defined as
the Fermi-surface average of the squared EP matrix elements for states at the Fermi level, divided by the eDOS at the Fermi energy:

\begin{equation}
  \begin{aligned}
    \teta_\k &=
    \frac{\Omega^2}{(2 \pi)^6 N(\vef)}
    \sum_{ij}^{N_{\text{states}}}
    \int_{\text{BZ}}
    d \bk
    \delta\left(\ve_{i, \bk} - \vef\right)
    \\
    &
    \times
    \int_{\text{BZ}}
    d \bkp
    \delta\left(\ve_{j, \bkp} - \vef\right)
    \\
    &
    \times
    \left|
    \bra{\Psi_{j, \bkp}}
    \bm{\nabla} V_\k(r)
    \ket{\Psi_{i, \bk}}
    \right|^2,
    \label{eq:eta_kappa_def}
  \end{aligned}
\end{equation}
where $N(\vef)$ is the eDOS at the Fermi level
\textit{per cell per spin};
$\vef$ is Fermi energy;
$\ve_{i, \bk}$ and
$\ve_{j, \bkp}$
are the Hamiltonian
eigenvalues of the states $i$ and $j$ corresponding to the
wavefunctions
$\Psi_{i, \bk}(\br)$ and $\Psi_{j, \bkp}(\br)$
at momenta $\bk$ and $\bkp$, respectively;
$N_{\text{states}}$ is the total number of electronic states;
$\Omega$ is the unit cell volume;
and ``BZ'' marks the integration over the Brillouin zone.
The eDOS in the denominator of Eq.~\eqref{eq:eta_kappa_def}
makes the atomic MH parameter and the corresponding atomic
EPC constant
dependent on the cell size.
The total EPC constant, $\lambda$ in Eq.~\eqref{eq:lambda_kappa}, however,
is independent of the cell size.

In materials where different Wyckoff positions are occupied,
it is convenient
to introduce the EPC constant $\lam_\mu$
for each of the $N_{\text{types}}$ symmetry types,
as defined in Eq.~\eqref{eq:lambda_mu}.
Each symmetry type includes only
atoms of the same chemical element that are related by
crystallographic symmetry operations,
with
$N_{\text{atoms}}^\mu$ being the number of atoms per type $\mu$.
The second equality of Eq.~\eqref{eq:lambda_mu} follows from
the fact that all atoms belonging to the same type $\mu$ share identical values of $\tilde{\lambda}_\k$.
\begin{equation}
  \begin{aligned}
  \lam_\mu &=
  \sum_{\k \in \{\mu\}}^{N_{\text{atoms}}^\mu} \tlam_\k
  =
  N_{\text{atoms}}^\mu
  \tlam_\k \Bigg|_{\forall \k \in \{\mu\}}.\\
  \label{eq:lambda_mu}
  \end{aligned}
\end{equation}

Unlike the atomic EPC constants, the type-specific
EPC constants, $\lam_\mu$,
are independent of the cell size.
The total EPC constant is then given by a
sum over the EPC constants corresponding to different symmetry types
[see Eq.~\eqref{eq:lambda_mu2}]
\cite{Pickett1982, Mazin1988, Quan2019, Hutcheon2020, Quan2021}.
Importantly, the ability to find the contributions of individual symmetry types to the overall
EPC in compounds
allows for a straightforward comparison between these contributions
in different materials.
\begin{equation}
  \begin{aligned}
  \lambda &=
  \sum_{\mu}^{N_{\text{types}}}
  \lam_\mu.
  \label{eq:lambda_mu2}
  \end{aligned}
\end{equation}

Since the masses $M_\k$ and the averaged phonon frequencies
$\left<\omega_\k^2\right>$ are identical for atoms of the same symmetry type,
MH parameters for different symmetry types can also be
introduced as
\begin{equation}
  \eta_\mu =
  N_{\text{atoms}}^\mu
  \teta_\k
  ~\forall~\k \in \{\mu\}.
  \label{eq:eta_mu}
\end{equation}
Similar to the type-specific EPC constants, the type-specific MH parameters
do not depend on the cell size.
Although the present formalism uses
atomic parameters,
only the type-specific ones
are reported in Section~\ref{sec:results} since they can be compared across different materials regardless of the unit cell choice.

\subsection{Assumption 3: Spherical-band approximation}

The RMTA assumes that the wavefunction for the state $i$
inside the MT sphere of atom $\k$, with radius $\rmt$,
can be
expanded
in terms of radial functions $R_{\k, l}(r, \ve_{i, \bk})$
corresponding to the local screened potential
$V_\k(r)$ and
spherical
harmonics $Y_{lm}(\hat{\br})$
characterized by quantum numbers $l$ and $m$,
with
momentum-dependent coefficients
$a_{i \k, lm}(\bk)$:

\begin{equation}
  \begin{aligned}
       \Psi_{i, \bk}(\br) &=
        \sum_{lm}
        a_{i \k, lm}(\bk)
        R_{\k, l}(r, \ve_{i, \bk})
        Y_{lm}(\hat{\br}).
  \end{aligned}
  \label{eq:rmta_wf_expansion}
\end{equation}

As shown in Appendix~\ref{app:ggp}, the
spherical-band approximation
that separates the radial ($k$) and angular ($\hat{\bk}$)
dependencies on the wavevector $\bk$
as
\begin{equation}
  a_{i \k, lm}(\bk) = a_{i \k, l}(k) Y^\ast_{lm}(\hat{\bk}),
  \label{eq:spherical_band}
\end{equation}
along with the separation
of radial ($r$) and angular ($\hat{\br}$)
components of $\br$,
is crucial. This approximation enables a significant simplification
of the MH parameter $\teta_\k$, reducing
Eq.~\eqref{eq:eta_kappa_def} to
\begin{equation}
  \begin{aligned}
    \teta_\k &=
    \sum_l
    \frac{2 (l+1)}{(2l+1)(2l+3)} M_{\k; l, l+1}^2
    \\
    &
    \times
    \frac{n_{\k, l}(\rmt, \vef)
    \times
    n_{\k, l+1}(\rmt, \vef)}
    {N(\vef)},
  \end{aligned}
  \label{eq:eta_ggp}
\end{equation}
where the partial EP matrix elements
are defined as
an integral of two normalized radial functions
with the potential gradient over the MT-sphere volume,
\begin{equation}
  \begin{aligned}
    &M_{\k; l, l+1}(\rmt, \vef)
    =
    \\
    &
    \frac{
    \int_0^{\rmt}
    dr r^2
    R_{\k, l}(r, \vef)
    \frac{d V_\k(r)}{d r}
    R_{\k, l + 1}(r, \vef)
    }
    {
    \sqrt{
    \int_0^{\rmt}
    d r r^2
    R^2_{\k, l}(r, \vef)
    \int_0^{\rmt}
    d r r^2
    R^2_{\k, l + 1}(r, \vef)
    }
    }.
  \end{aligned}
  \label{eq:mll1_def}
\end{equation}

As a result of the spherical-band approximation,
Eq.~\eqref{eq:eta_ggp}
includes only partial EP matrix elements
corresponding to
dipole transitions with $\Delta l = \pm 1$. The quantities $n_{\k, l}$ represent the partial eDOS
for angular momentum $l$
inside the MT sphere of
atom $\k$, defined as
\begin{equation}
  \begin{aligned}
    n_{\k, l}&(\rmt;
    \vef)
    =
    \sum_m n_{\k, lm}(\rmt;
    \vef)
    \\
    &
    =
    \sum_m
    \frac{\Omega}{(2 \pi)^3}
    \int_0^{\rmt}
    d r r^2 R^2_{\k, l}(r, \vef)
    \\
    &
    \times
    \sum_i
    \int_{\text{BZ}} d \bk
    \delta(\ve_{i, \bk} - \vef)
    \left|a_{i \k, lm}(\bk)\right|^2,
  \end{aligned}
  \label{eq:nl_def}
\end{equation}
It is important to point out
that the
spherical-band approximation is not required for
the partial eDOS calculation,
as can be seen in Eq.~\eqref{eq:nl_def},
where general coefficients
$a_{i\k,lm}(\bk)$ are used.

Hereafter, the task
of computing MH parameters
in PS+PW codes
comes down to the calculation
of partial EP matrix elements
and partial eDOS.

\subsection{Partial EP matrix elements evaluated
on the MT-sphere}
\begin{figure}[!t]
  \includegraphics[width=0.95\columnwidth]{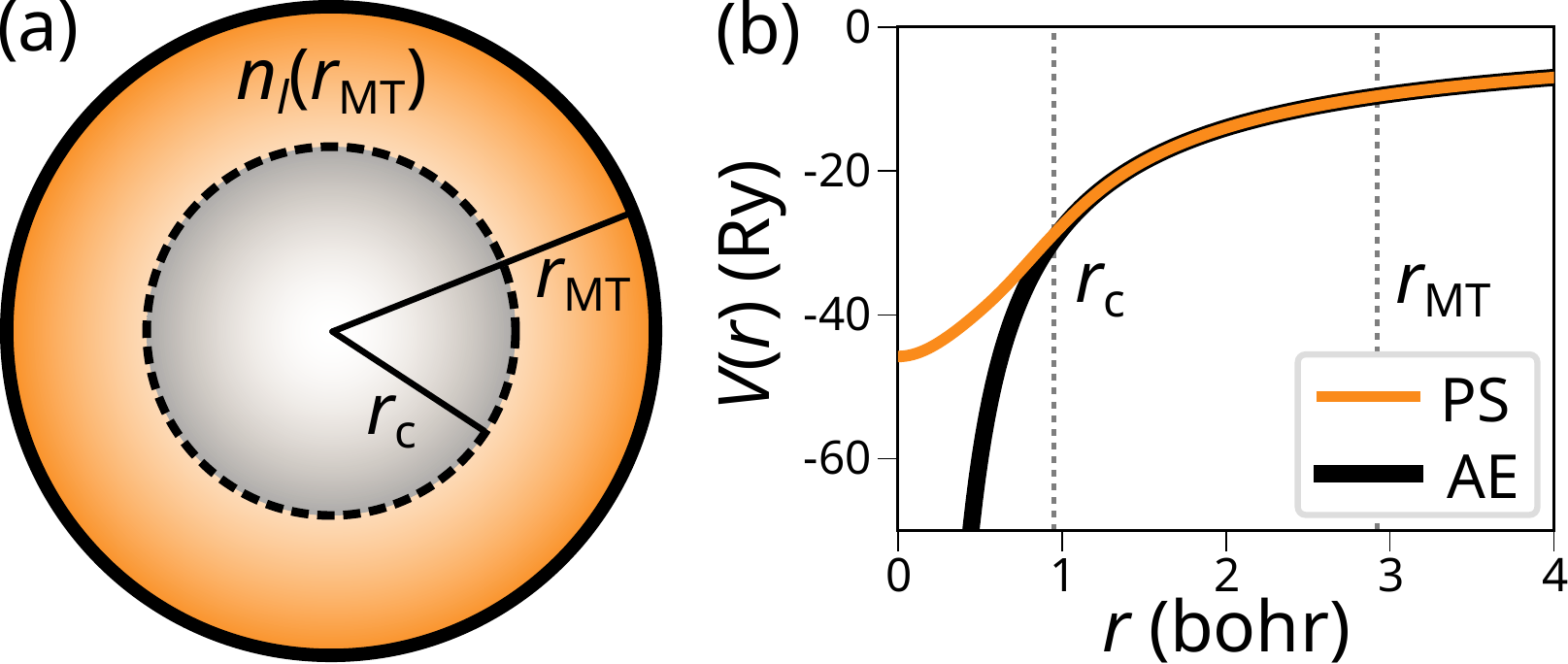}
  \caption{(a) Muffin-tin radius, $r_{\text{MT}}$,
  compared to pseudopotential cut-off
  radius, $r_{\text{c}} < r_{\text{MT}}$;
  (b) difference between
  pseudopotential (PS)
  and all-electron (AE) potential
  for radii below $r_{\text{c}}$.
  }
  \label{fig:ps_vs_ae}
\end{figure}

The original expression for the
partial EP matrix elements, introduced
by Gaspari and Gyorffy [Eq.~\eqref{eq:mll1_def}] \cite{GaspariGyorffy1972, Evans1973},
involves integration
of the radial functions $R_{\k, l}(r, \vef)$
and $R_{\k, l + 1}(r, \vef)$,
along with the derivative of
the MT potential $d V_\k(r) / d r$,
over the volume of the MT-sphere [Fig.~\ref{fig:ps_vs_ae}(a)].
This expression has three problems for the
PS+PW codes:
(1)~A pseudopotential (PS) differs from the
all-electron (AE) potential within
a cut-off radius $r_{\text{c}}$,
where pseudization takes place,
rendering the integration
of the PS derivative in that region
unreliable;
(2)~The screened self-consistent potential,
as calculated in PS+PW codes,
is periodic in the real space,
and needs to be converted to localized
screened atomic potentials;
(3)~The radial functions $R_{\k, l}(r, \ve)$
are not calculated explicitly and need to
be derived from available DFT data.

In our method, the first problem is solved
by referring to the following reformulated
expression for Eq.~\eqref{eq:mll1_def}:

\begin{equation}
  \begin{aligned}
    &M_{\k; l, l+1}(r_{\text{MT}}, \vef) =\\
    &
    \frac{[V_{\k}(r) - \vef]r^2 -
    (L_{\k, l} - l)[L_{\k, l+1} + (l + 2)]}
    {r \sqrt{\frac{\partial L_{\k, l}}{\partial \ve} \cdot
    \frac{\partial L_{\k, l+1}}{\partial \ve}}}
    \Bigg|_{
    \substack{
    \ve = \vef\\
    r = \rmt
    }
    },
    \\
    &
    L_{\k, l}(r, \ve) = r
    \frac{
    \left(
    \frac{\partial R_{\k, l}(r, \ve)}
    {\partial r}
    \right)
    }
    {R_{\k, l}(r, \ve)}.
  \end{aligned}
  \label{eq:mll1_pet}
\end{equation}
It was
introduced by Pettifor \cite{Pettifor1977}
based on the radial Schr\"{o}dinger equation
for radial functions $R_{\k, l}(r, \ve)$
and their normalization conditions, provided in Appendix~\ref{app:norm_to_dlde}.
Here,
$V_{\k}(r)$ is the local screened potential
of the atom $\k$,
$L_{\k, l}(r, \ve)$
is the logarithmic derivative of the
radial part
of the
wavefunction evaluated around the Fermi level,
and $\frac{\partial L_{\k, l+1}}{\partial \ve}$
is the energy derivative of the logarithmic derivative.
In this expression,
the values of the potential and
logarithmic derivatives at the MT-radius
are sufficient to calculate the partial EP matrix elements.

\begin{figure}[!t]
  \includegraphics[width=0.9\columnwidth]{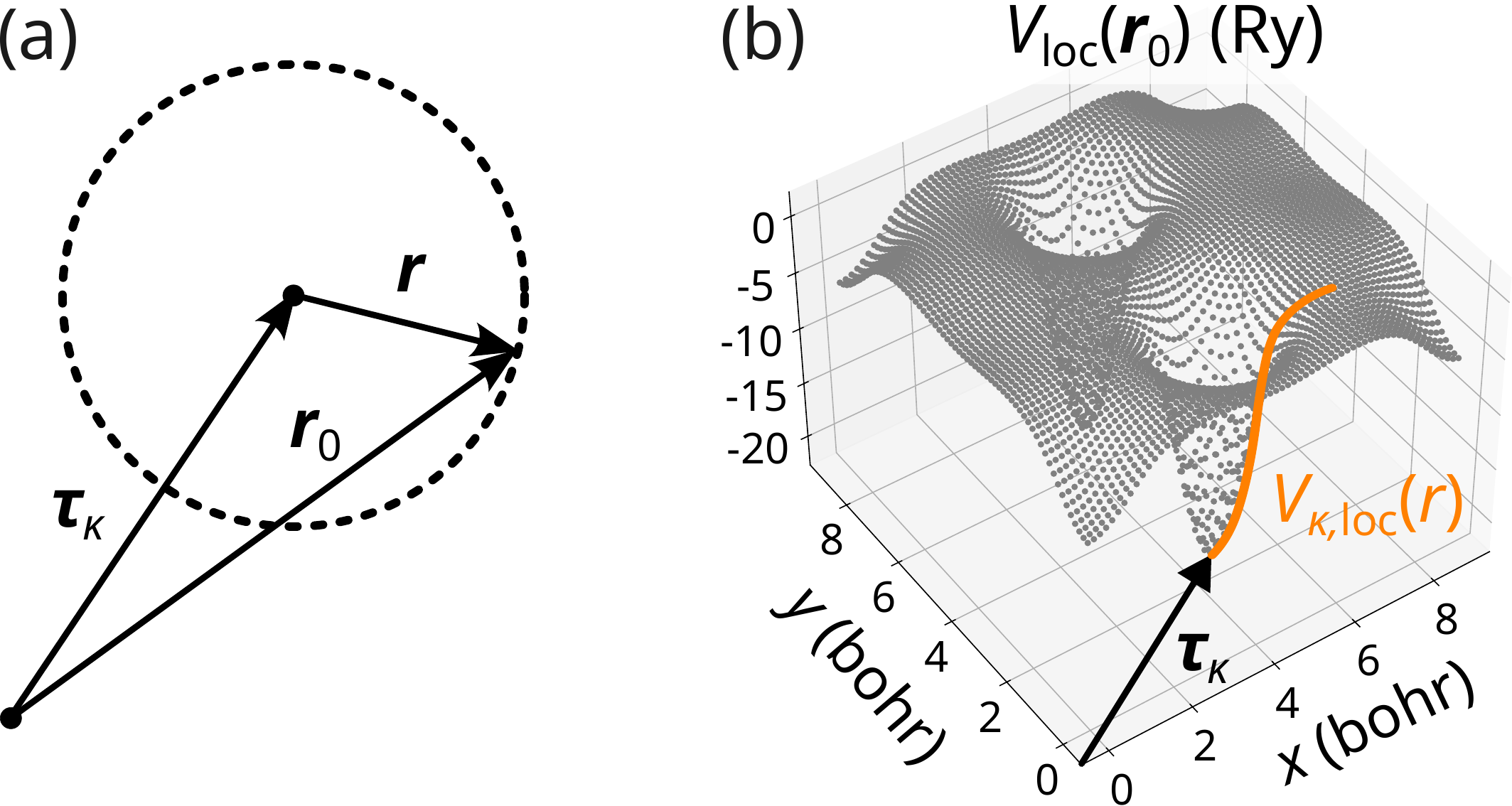}
  \caption{(a) Relation between real-space grid $\br$,
  centered on the atom at $\bm{\tau}_\k$,
  and grid $\br_0$,
  centered on the origin of the calculated cell;
  (b) spherically-symmetric atomic local
  potential
  $V_{\k, \text{loc}}(r)$ extracted
  from periodic local
  potential with Eq.~\eqref{eq:vkappa}.
  }
  \label{fig:extract_vkappa}
\end{figure}

The second problem is addressed by
using properties of the Fourier transformation.
The self-consistent DFT pseudopotential is given as
a periodic local screened potential
$V_{\text{loc}}(\br_0)$
[Fig.~\ref{fig:extract_vkappa}(b)]
on the 3D real-space grid $\br_0$,
with a center
at an arbitrary origin of the calculated unit cell.
The Fourier transform of this potential is represented by
coefficients
\begin{equation}
  \begin{aligned}
    \tilde{V}(\bG) &=
    \frac{1}{\Omega}
    \int_{\Omega} d \br_0
    e^{- i \bG \cdot \br_0}
    V_{\text{loc}}(\br_0),
  \end{aligned}
  \label{eq:vg}
\end{equation}
where $\bG$ is a reciprocal lattice vector.

A spherically-symmetric local screened atomic potential, $V_{\k, \text{loc}}(r)$,
is then
expressed in Eq.~\eqref{eq:vkappa} by
obtaining coefficients of the plane-wave
expansion of the periodic potential on a shifted grid, $\br$,
that is centered
on a specific atom at $\bm{\tau}_\k$
[Fig.~\ref{fig:extract_vkappa}(a-b)],
as well as by using the
expansion of $e^{i \bG \cdot \br}$
in spherical Bessel functions and
spherical harmonics
for only $l = m = 0$
[see Appendix~\ref{app:vkappa} for details].
\begin{equation}
  \begin{aligned}
    V_{\k, \text{loc}}(r) &=
    V_{\k, \text{loc}}(\left|\br_0 - \bm{\tau}_\k \right|)
    \\
    &=
    \sum_{\bG}
    e^{i \bG \cdot \bm{\tau}_\k}
    \tilde{V}(\bG)
    \frac{\sin{G r}}{G r}.
  \end{aligned}
  \label{eq:vkappa}
\end{equation}
In this equation,
$r~=~|\br|$, $G~=~|\bG|$, and
$e^{i \bG \cdot \bm{\tau}_\k}\tilde{V}(\bG)$
are modified coefficients of the
plane-wave expansion of the same
periodic potential
$V_{\text{loc}}(\br)$
[Fig.~\ref{fig:extract_vkappa}(b)]
but
on a shifted radial grid, $\br$,
that is centered on the atom.
The potential
$V_{\text{loc}}(\br)$
is DFT-screened.
Hence, the extracted local
pseudopotential
$V_{\k, \text{loc}}(r)$
is screened, too.
Note that the local PS is
equal to the AE potential above the cut-off radius:
$V_{\k}(r) = V_{\k, \text{loc}}(r)~\forall~r > r_{\text{c}}$.

\begin{figure}[!t]
  \includegraphics[width=1\columnwidth]{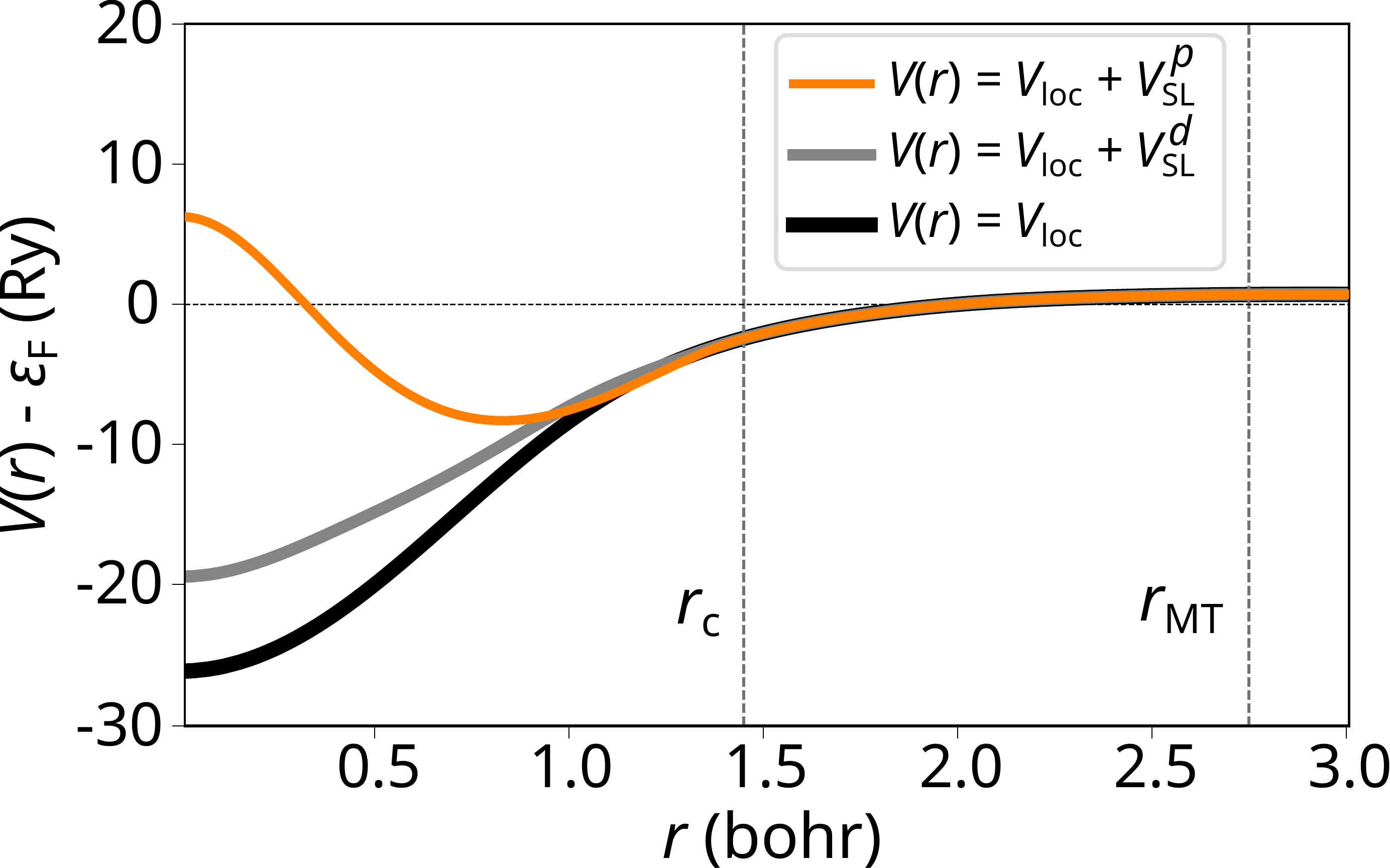}
  \caption{
  Effect of semilocal, $l$-dependent,
  contributions added to the
  local
  pseudopotential of Nb atom
  in Nb BCC structure.
  }
  \label{fig:semilocal}
\end{figure}

For the third problem,
the method of choice is to generate
radial functions $R_{\k, l}(r, \ve)$,
which are
compatible with the
extracted screened potential
$V_{\k, \text{loc}}(r)$,
for a particular energy $\ve$
by
solving the corresponding radial Schr\"{o}dinger
equation,
\begin{equation}
  \begin{aligned}
    &\left[
    - \frac{d^2}{d r^2}
    +
    V^l_{\k, \text{eff}}(r)
    \right]
    \left[r R_{\k, l}(r, \ve)\right]
    =
    \ve \left[r R_{\k, l}(r, \ve)\right]
    \Big|_{\ve = \vef},
  \end{aligned}
  \label{eq:rad_schr}
\end{equation}
with an effective potential $V^l_{\k, \text{eff}}(r)$
containing $V_{\k, \text{loc}}(r)$,
$l$-dependent semilocal (SL) potential $V^{l}_{\k, \text{SL}}(r)$
that is non-zero only in the core region $r < r_{\text{c}}$ (see Fig.~\ref{fig:semilocal}),
and centrifugal potential $l (l+1) / r^2$:

\begin{equation}
  \begin{aligned}
    V^l_{\k, \text{eff}}(r) &=
    V_{\k}(r) + \frac{l (l+1)}{r^2}\\
    &=
    \left[
    V_{\k, \text{loc}}(r) + V^{l}_{\k, \text{SL}}(r)
    \right]
    +
    \frac{l (l+1)}{r^2}.
  \end{aligned}
  \label{eq:veff}
\end{equation}

Eq.~\eqref{eq:rad_schr}
is solved numerically at the specified energy $\ve$ by outward integration from $r$ value close to zero to the desired MT-radius.
This way, the energy derivatives of the
logarithmic derivatives $\partial L_{\k, l} / \partial \ve$
can be obtained by solving the radial Schr\"{o}dinger
equation at different energies close to the Fermi level.
The contribution from the SL potential is pertinent to the PS approach:
Different pseudopotentials are
needed to generate radial functions for different
$l$ values, while only
one AE potential would generate all radial functions.

Although most PS+PW codes use pseudopotentials
in the non-local (NL) (dependent on $r$ and $r^\prime$)
Vanderbilt-Kleinman-Bylander (VKB) form
\cite{Kleinman1982, Bylander1990, Vanderbilt1990}
due to its computational efficiency in reciprocal space,
we use the SL form for simplicity in the current
implementation.
For the non-local ONCVPSP pseudopotentials by Hamann
\cite{Hamann1979, Hamann2013}
utilized in this work, automatic
conversion of the NL components, read directly from
the pseudopotential file, to the SL form
is straightforward and described in Appendix~\ref{app:hamann}.
In principle, solving the
radial Schr\"{o}dinger
equation with NL potentials directly \cite{Gonze1991}
would give the same results.

\subsection{Partial eDOS inside the MT-sphere and total eDOS}
The partial eDOS inside the MT-sphere of atom $\k$
are derived based on the assumption that
on the surface of the MT-sphere
the wavefunction of the state $i$
can be expanded in both spherical harmonics
and plane waves at the same time:

\begin{equation}
    \begin{cases}
       \Psi_{i, \bk}(\br) &=
        \sum_{lm}
        a_{\k i, lm}(\bk)
        R_{\k, l}(\rmt, \ve_{i, \bk})
        Y_{lm}(\hat{\br}),\\
       \Psi_{i, \bk}(\br) &=
       \psi_{i, \bk}(\bm{\tau}_\k + \br)
       \Big|_{r = \rmt}
       \\
       &=
        \frac{1}{\sqrt{\Omega}}
        \sum_{\bG}
        e^{i (\bk + \bG) \cdot
        \left( \bm{\tau}_\k + \br \right)}
        \tilde{\psi}_i(\bk + \bG)
        \Bigg|_{r = \rmt},
    \end{cases}
    \label{eq:double_expansion}
\end{equation}
where
$\tilde{\psi}_i(\bk + \bG)$ are
coefficients of the Fourier expansion
of the wavefunction $\Psi_{i, \bk}(\br)$.

From the system of Eqs.~\eqref{eq:double_expansion}, the coefficients $a_{\k i, lm}(\bk)$
are found following Eq.~\eqref{eq:a_of_k},
which is calculated by using the Gauss-Legendre quadrature method \cite{Rabinowitz2007}.

\begin{equation}
    \begin{aligned}
       a&_{\k i, lm}(\bk)
       =
       \frac{1}{\sqrt{\Omega} R_{\k, l}(r, \ve_{i, \bk})}
       \\
       &\times
       \int d \hat{\br}
       \left[
       \sum_{\bG}
        e^{i (\bk + \bG) \cdot
        \left( \bm{\tau}_\k + \br \right)}
        \tilde{\psi}_i(\bk + \bG)
       \right]
       Y^\ast_{lm}({\hat{\br}})
       \Bigg|_{r = \rmt}.
    \end{aligned}
    \label{eq:a_of_k}
\end{equation}

Then, using Eqs.~\eqref{eq:a_of_k} and
\eqref{eq:nl_def_simple} (see Appendix \ref{app:ggp}),
the partial eDOS, $n_{\k, l}$, can be evaluated
from the Fourier coefficients,
$\tilde{\psi}_i(\bk + \bG)$,
and the logarithmic derivatives,
$\frac{\partial}{\partial \ve} L_{\k, l}$,
by Eq.~\eqref{eq:nl}.
Note that this method
allows for the calculation of not only the $l$-resolved eDOS,
but also $m$-resolved eDOS, $n_{\k, lm}$.

\begin{equation}
    \begin{aligned}
       &n_{\k, l}(\rmt, \vef)
       =
       \sum_{m = -l}^{l}
       n_{\k, lm}(\rmt, \vef)
       \\
       &=
       \sum_{m = -l}^{l}
       \frac{\rmt}{(2 \pi)^3}
       \left(
       -
       \frac{\partial}{\partial \ve} L_{\k, l}
       \right)
       \int_{\text{BZ}} d \bk
       \sum_{i = 1}^{N_{\text{states}}}
       \delta(\ve_{i, \bk} - \vef)
       \times\\
       &
       \left|
       \int d \hat{\br}
       \left[
       \sum_{\bG}
       e^{i (\bk + \bG) \cdot (\bm{\tau}_\k + \br)}
       \tilde{\psi}_i(\bk + \bG)
       \right]
       Y^\ast_{lm}({\hat{\br}})
       \right|^2
       \Bigg|_
       {\substack{
       r = \rmt\\
       \ve = \vef
       }}.
    \end{aligned}
    \label{eq:nl}
\end{equation}

As can be seen from Eq.~\eqref{eq:mll1_pet}, the squared partial EP matrix element, $M^2_{l, l+1}$,
contains energy derivatives of the logarithmic derivatives,
$\frac{\partial}{\partial \ve} L_{\k, l}$ and $\frac{\partial}{\partial \ve} L_{\k, l + 1}$,
in the denominator. These energy derivatives are normalization integrals for the radial functions
$R_l$ and $R_{l+1}$ (see Appendix~\ref{app:norm_to_dlde})
that cancel out when $M^2_{l, l+1}$ is multiplied by the partial
eDOS $n_l$ and $n_{l+1}$ from Eq.~\eqref{eq:nl}.
Hence, if the calculation of the partial eDOS themselves
is not required, Eq.~\eqref{eq:eta_ggp} for the MH parameters can be rewritten in a form independent
of $\frac{\partial}{\partial \ve} L_{\k, l}$ and $\frac{\partial}{\partial \ve} L_{\k, l + 1}$,
such that energy derivatives of the logarithmic derivatives, or normalization integrals,
do not have to be computed.

The eDOS at the Fermi level per cell per spin
is computed as a Brillouin-zone integral of the delta-functions summed over all electronic states,

\begin{equation}
  \begin{aligned}
    N(\vef) =
    \frac{\Omega}{(2 \pi)^3}
    \int_{\text{BZ}} d \bk
    \sum_i \delta \left( \ve_{i, \bk} - \vef\right).
  \end{aligned}
  \label{eq:}
\end{equation}

\section{\label{sec:details}Technical details}

The present implementation was developed within the \textsc{Quantum ESPRESSO} (QE)
package
\cite{Giannozzi2009, Giannozzi2017, Giannozzi2020}, which was employed in this study for calculating the self-consistent
pseudopotentials with the plane-wave (PW) code.
Norm-conserving ONCVPSP \cite{Hamann1979, Hamann2013} pseudopotentials
with the Perdew-Burke-Ernzerhof (PBE) \cite{Perdew1996}
generalized-gradient approximation (GGA) for the exchange-correlation functional
were used for all atoms,
with a wavefunction energy cutoff of 120~Ry.
The RMTA code was executed as a postprocessing step
after the self-consistent field (SCF) calculation.
For the quantitative verification presented in Sec.~\ref{sec:results},
all structures were modeled with their primitive unit cells, and
MT-radii were chosen to correspond to touching spheres.
The default three-dimensional FFT mesh for charge density and self-consistent potential, automatically determined by the energy cutoff, was retained.
A $24 \times 24 \times 24$ Monkhorst-Pack $\bk$-point grid
and
Methfessel-Paxton smearing \cite{Methfessel1989} with $\sigma = 10^{-2}$~Ry
were utilized
for the approximation of the occupation function in
the SCF calculation.
The tetrahedron method of integration \cite{Blochl1994}
on the same $\bk$-point grid was selected for
the evaluation of the partial and total eDOS
across all structures,
since it showed faster convergence
with respect to the $\bm{k}$-point grid
than
smeared delta-function approximations (see Section II of the Supplementary Information \cite{Supplemental}).

Calculated partial EP matrix elements and
McMillan-Hopfield parameters were compared
to the FLAPW code \textit{flair} \cite{Weinert2009}.
In FLAPW, semicore states
of all atoms were treated as core states and included explicitly in core-orthogonalization.
Equations~\eqref{eq:eta_ggp} and \eqref{eq:mll1_pet}
were evaluated with the atomic potential, radial functions,
and
partial and total eDOS calculated using the FLAPW method.
The same
MT-radii,
$\bk$-point grid, and smearing parameters as those used for the RMTA calculations with the PW code
were employed to ensure consistency.

\section{\label{sec:results}Quantitative verification}
\begin{figure}[!t]
  \includegraphics[width=1\columnwidth]{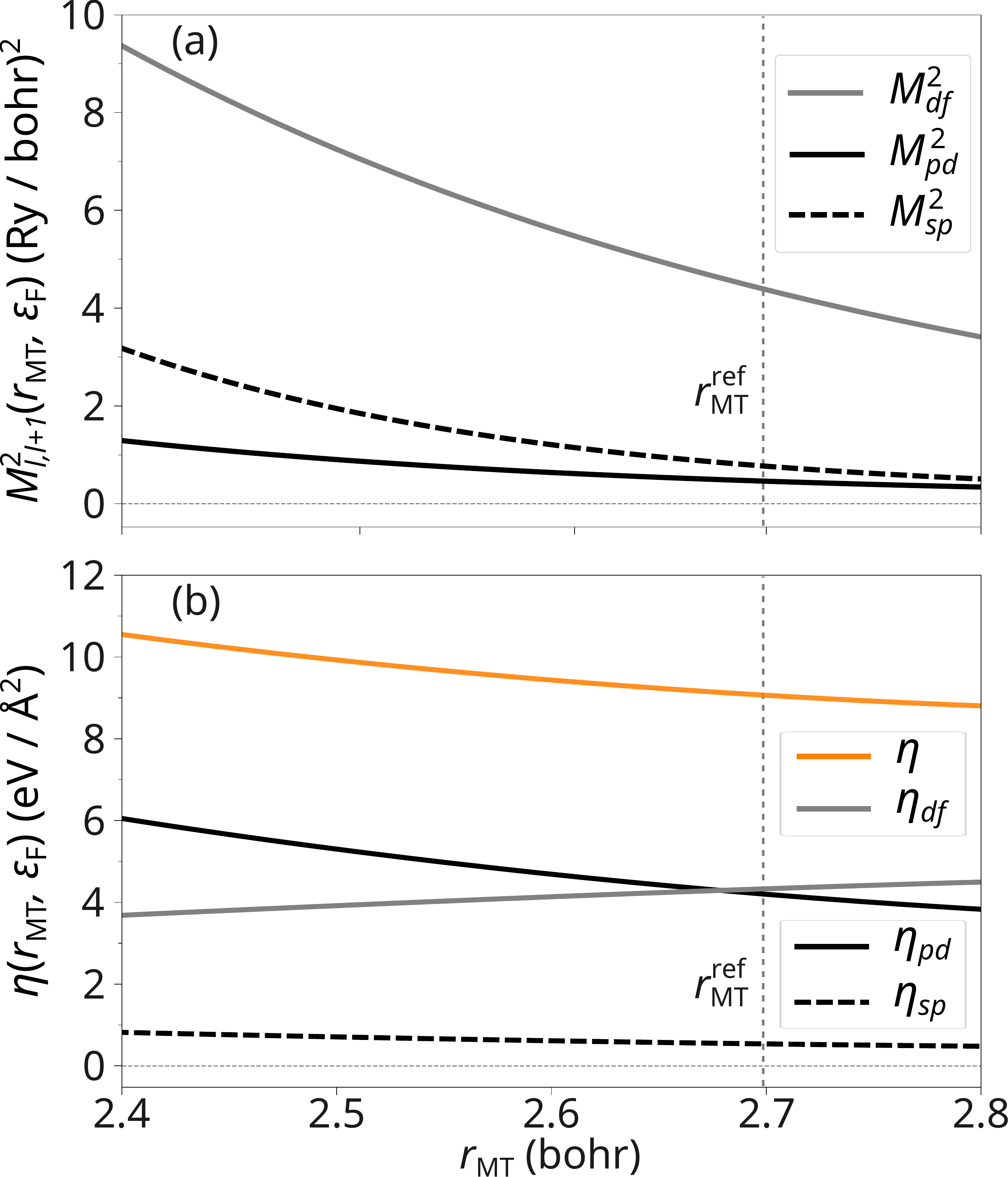}
  \caption{(a) Partial EP
  matrix elements, $M_{l, l+1}^2$, and
  (b) partial, $\eta_{l, l+1}$, and
  total, $\eta$, MH  parameters
  as functions of the muffin-tin
  radius, $r_{\text{MT}}$,
  for the Nb atom in the BCC structure.
  Dashed vertical lines show the
  reference muffin-tin radius
  at $r_{\text{MT}}^{\text{ref}} = 2.70$~bohr.
  }
  \label{fig:m2_and_eta_in_nb_bcc}
\end{figure}

A key difference in how the partial EP matrix
elements and MH parameters depend on the choice of the MT-radius is illustrated in
Figs.~\ref{fig:m2_and_eta_in_nb_bcc}(a)-(b) for the BCC structure of Nb.
For all $l$ channels, $M^2_{l, l+1}$ decrease as MT-radius increases [Fig.~\ref{fig:m2_and_eta_in_nb_bcc}(a)], consistent with Pettifor's observation \cite{Pettifor1977} that partial EP matrix elements alone
are not sufficient to describe the electronic part of the EPC constant because they are strongly dependent on
the MT-radius.

On the contrary, the partial MH parameters,
which are presented in eV / \AA$^2$
following previous studies
\cite{Pettifor1977, Papaconstantopoulos1977, Pickett1982},
appear to be less sensitive [Fig.~\ref{fig:m2_and_eta_in_nb_bcc}(b)] to the choice
of the MT-radius since they
include the partial eDOS, which increases as the MT sphere expands. Notably, while the
$df$ channel may seem dominant based on its corresponding partial EP matrix element, it contributes roughly the same to the total MH parameter
$\eta$ as the $pd$ channel at the reference MT radius.
Achieving a nearly-flat behavior of the total atomic MH parameter
around the reference MT-radius was one of the objectives
of this work. Nevertheless, the slight dependence on $r_{\text{MT}}$ indicates that careful selection of the MT-radius based on
the cell geometry is advisable.

\begin{figure}[!t]
  \includegraphics[width=1\columnwidth]{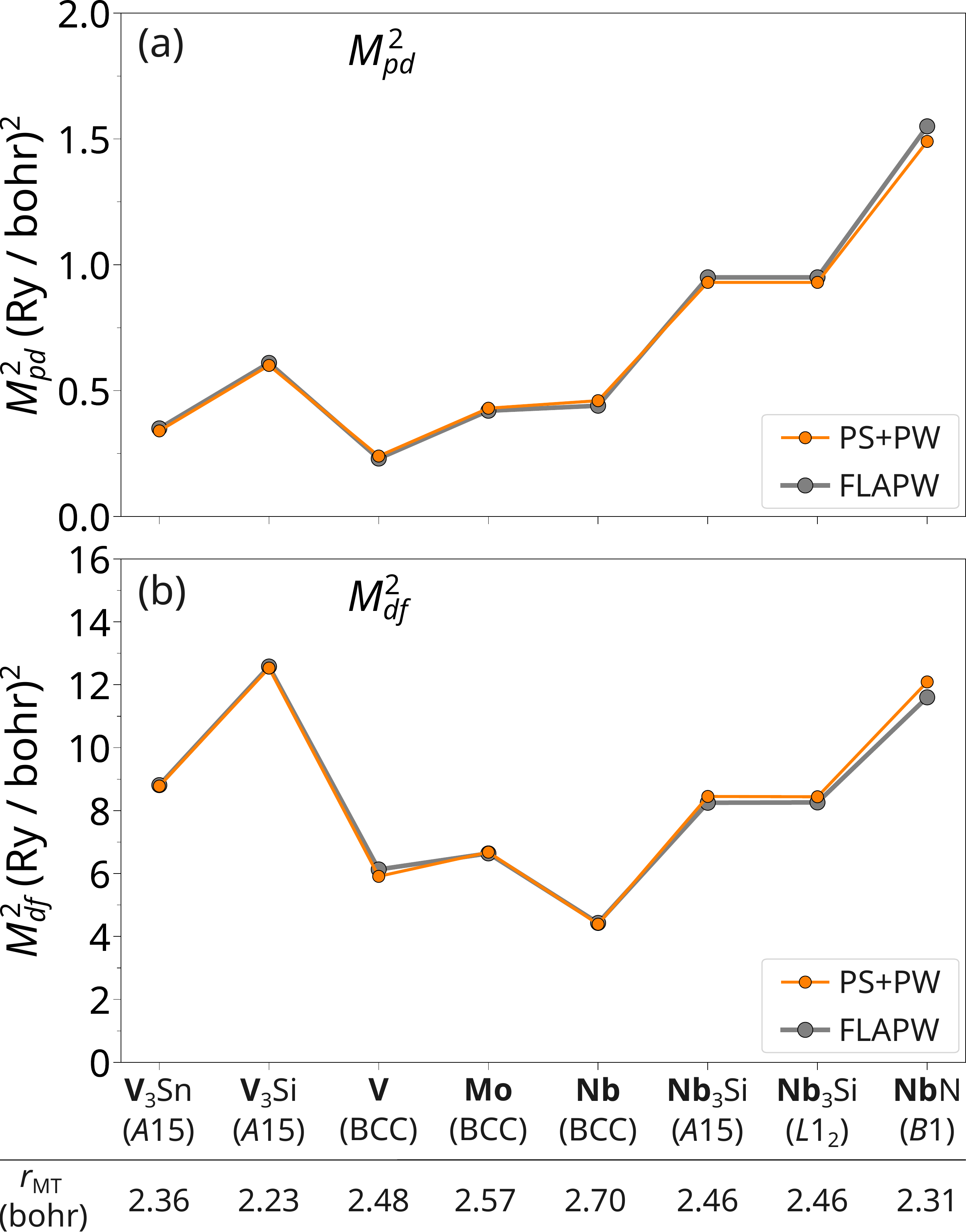}
  \caption{
  Comparison of squared partial EP
  matrix elements, $M^2_{l, l+1}$, between PS+PW and FLAPW
  methods for
  metal atoms--V, Mo, and Nb--in different structures:
  (a) $pd$ channel
  and (b) $df$ channel.
  Corresponding MT-radii of the metal atoms
  are listed at the bottom of the figure.
  }
  \label{fig:compare_to_flapw_m2}
\end{figure}

\begin{figure}[!t]
  \includegraphics[width=0.98\columnwidth]{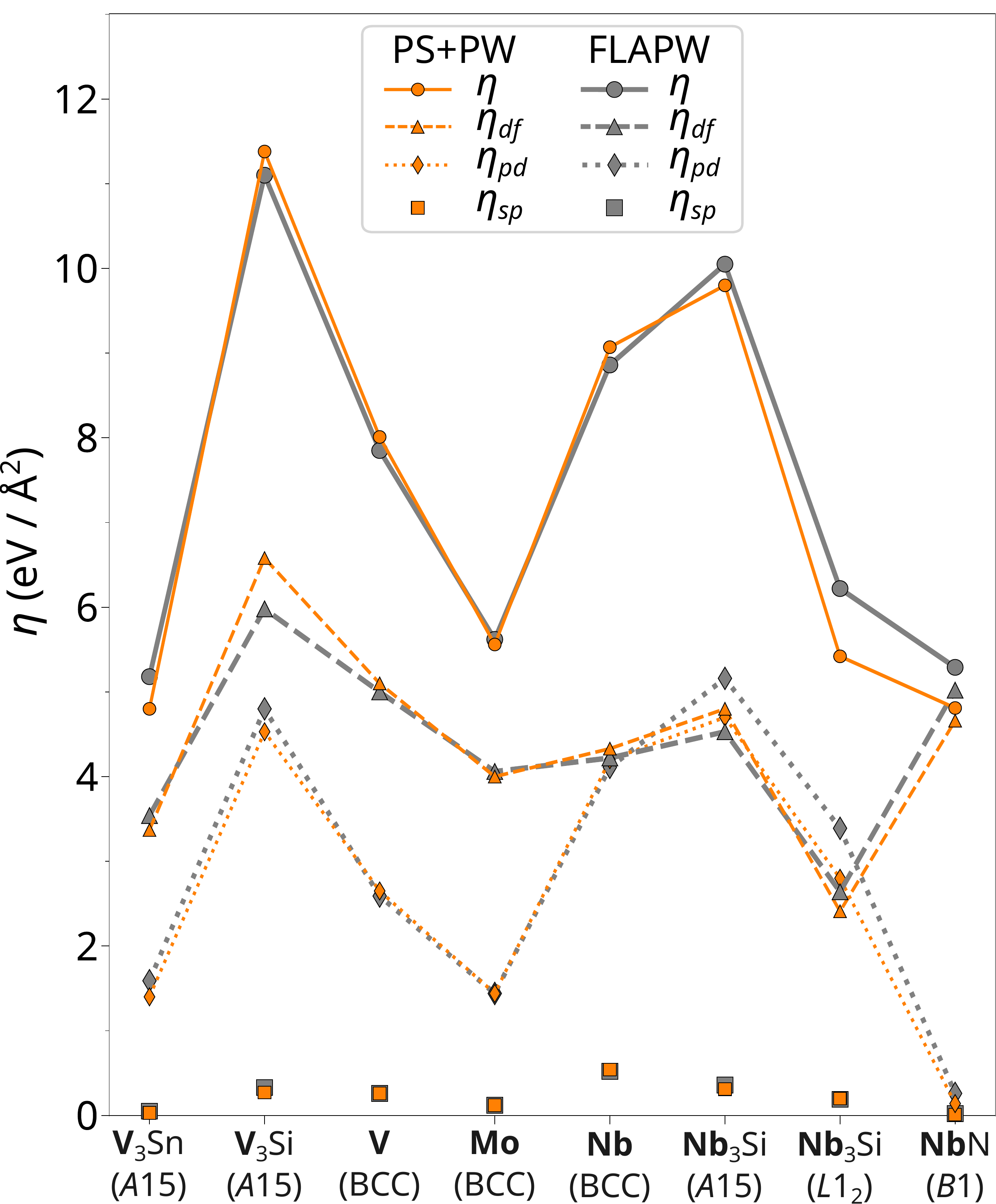}
  \caption{
  Comparison of partial, $\eta_{l, l+1}$,
  and total, $\eta$,
  type-resolved MH parameters
  between PS+PW and FLAPW
  methods for
  types containing only
  metal atoms--V, Mo, and Nb--in different structures.
  }
  \label{fig:compare_to_flapw_eta}
\end{figure}

The good agreement between the PS+PW and FLAPW
methods for the electronic part
calculated in selected
simple metals and compounds is demonstrated
in Figs.~\ref{fig:compare_to_flapw_m2} and \ref{fig:compare_to_flapw_eta},
where
\textit{Strukturbericht}
symbols \cite{Hermann1931} are used as space-group
identifiers for compounds.
Crystallographic parameters of the structures are provided in
the Supplementary Information \cite{Supplemental}.
The differences in partial EP matrix
elements ($pd$ and $df$) for metal atoms in the selected
structures do not exceed 5\% [Fig.~\ref{fig:compare_to_flapw_m2}],
with PS+PW and FLAPW results almost overlapping.
In Fig.~\ref{fig:compare_to_flapw_eta}, which shows
MH parameters for symmetry types of the metal atoms, both partial and total MH parameters exhibit similar trends across the two methods.
Both methods demonstrate
that $\eta$ parameter of Nb atoms in
NbN is mostly dominated by the $df$ channel contribution,
in agreement
with previous studies \cite{Pickett1982, Klein1979}.
The observed discrepancies bewteen PS+PW and FLAPW
are primarily due to slight differences in the electronic density of states.
Importantly, there is no tendency for one of the methods to produce consistently higher
values than the other.
Comparison of the total MH parameters,
shown as circles in Fig.~\ref{fig:compare_to_flapw_eta},
to some previous APW results \cite{Papaconstantopoulos1977, Pickett1982}
is discussed in Section III of the Supplementary Information \cite{Supplemental}.

Table~\ref{tab:lambda} compares the calculated MH parameters for simple metals with the results of Papaconstantopoulos \textit{et al.} \cite{Papaconstantopoulos1977}, and uses them to estimate the EPC constant by approximating the “phonon” contribution via the experimental Debye temperature.
It should be noted that the MH parameters from Ref.~\cite{Papaconstantopoulos1977} were obtained with the
augmented plane wave (APW) method and Moruzzi-Janak-Williams (MJW) LDA
potentials \cite{Moruzzi1978}.
The noticeable differences from the present results are likely due to the evolution of
potentials and the use of different exchange-correlation functionals.
While the prior MH parameters tend to be lower than those obtained in this work, the
overall trend across elemental metals remains similar.
This consistency suggests that the RMTA can be effectively used
to identify materials with a higher likelihood of superconductivity,
provided that the same RMTA implementation is applied consistently across all candidates.

\begin{table}[htbp]
  \caption{\label{tab:lambda}Estimated semi-phenomenological EPC
  constant, $\lambda$, based on calculated
  McMillan-Hopfield parameters, $\eta$, at specified MT-radii, $r_{\text{MT}}$,
  and experimental Debye temperature,
  $\theta_{\text{D}}$, for simple metals.
  $\theta_{\text{D}}$ values are taken from corresponding references.
  MH parameters $\eta_{\text{ref}}$,
  calculated
  with the APW method and MJW potentials
  \cite{Papaconstantopoulos1977},
  are provided for the reference.}
  \begin{tabular}{cccccc}
    \hline\hline
    \\
    Structure
    &
    $r_{\text{MT}}$
    &
    $\eta$
    &
    $\eta_{\text{ref}}$ \cite{Papaconstantopoulos1977}
    &
    $\theta_{\text{D}}$
    &
    $\lambda$
    \\
    &
    (bohr)
    &
    $\left(\frac{\text{eV}}{\text{\AA{}}^2}\right)$
    &
    $\left(\frac{\text{eV}}{\text{\AA{}}^2}\right)$
    &
    (K)
    \\
    \\
    \hline
    \\
    V (BCC)  & 2.48 & 8.01 & 6.89 & 385 \cite{Desai1986} & 1.19\\
    Cu (FCC) & 2.41 & 0.57 & 0.46 & 343 \cite{Kittel2004} & 0.09\\
    Nb (BCC) & 2.70 & 9.07 & 7.63 & 275 \cite{Arblaster2017} & 1.45\\
    Mo (BCC) & 2.57 & 5.56 & 5.80 & 430 \cite{Desai1987} & 0.35\\
    Rh (FCC) & 2.56 & 4.95 & 4.72 & 480 \cite{AIP1972} & 0.24\\
    Pd (FCC) & 2.66 & 2.65 & 1.99 & 299 \cite{Rayne1957} & 0.31\\
    Ag (FCC) & 2.73 & 0.28 & 0.29 & 225 \cite{Kittel2004} & 0.06\\
    \\
    \hline\hline
  \end{tabular}
\end{table}

\section{\label{sec:discs}Discussions}

Machine learning studies \cite{Hutcheon2020, Shipley2021, Bai2024} have demonstrated that McMillan-Hopfield parameters and partial eDOS can serve as effective descriptors for identifying potential superconductors. In this context, the present method is well-suited for high-throughput screening of candidate materials. As a minimal criterion, materials with very low MH parameters can be reliably excluded from consideration, as they are unlikely to exhibit superconductivity.

For a more complete estimate of the EPC constant,
the phonon part, given by the averaged squared phonon frequencies
$\left< \omega^2_{\k} \right>$, still needs to
be evaluated using a computationally efficient approach. One practical strategy is to replace the direct evaluation of $\left< \omega^2_{\k} \right>$
with that of the
interatomic force constants. These can be
approximated from the diagonal elements of the inverted
dynamical matrix, obtained at a few high-symmetry points on the Brillouin zone boundary via phonon linear-response calculations \cite{Gomersall1974Aug, Dicks2024},
or from the diagonal elements of the inverse force-constant
matrix at the $\Gamma$-point in frozen-phonon calculation,
assuming a sufficiently large supercell
\cite{Foulkes1975, Ginzburg1982, Mazin1990}.

To improve RMTA predictions for $sp$-metals,
corrections
to the asymptotic behavior of the local ionic
potential
could be introduced, as done by
Zdetsis~\textit{et al.} \cite{Zdetsis1981}
and Mazin~\textit{et al.} \cite{Mazin1984}.

\section{Conclusions}
We presented a method for implementing the rigid
muffin-tin approximation
in pseudopotential-based plane-wave (PS+PW)
codes to calculate the electronic part of the EPC constant, expressed through the McMillan-Hopfield parameter.
The calculated values for a variety of simple metals and
compounds
were validated against results from a full-potential
augmented plane wave code.
The excellent agreement in partial EP
matrix elements and McMillan-Hopfield parameters
between the pseudopotential and all-electron methods confirms the reliability of the developed methodology. Some discrepancies with older references
likely reflect the evolution of potentials and
exchange-correlation functionals over time.
As highlighted in previous machine learning studies,
McMillan-Hopfield parameters computed within the PS+PW framework
can be used as descriptors in machine learning algorithms for high-throughput
screening of potential superconducting materials.

\begin{acknowledgments}
This work was supported by National Science Foundation (Grants No. DMR-2320074,  DMR-2320073, and DMREF-2323857).
\end{acknowledgments}


\appendix

\section{\label{app:ggp}Assumptions of spherical potential and bands for McMillan-Hopfield parameters}
This
section demonstrates how the assumptions of a
spherical potential and
spherical bands
lead to simplified
expressions for the MH
parameters.
Following
\cite{GaspariGyorffy1972, Evans1973, Pettifor1977},
Eq.~\eqref{eq:eta_kappa_def} can be rewritten as
\begin{equation}
  \begin{aligned}
    \teta_\k &=
    \frac{\Omega^2}{(2 \pi)^6
    N(\vef)
    }
    \sum_{ij}^{N_{\text{states}}}
    \int_{\text{BZ}}
    d \bk
    \delta\left(\ve_{i, \bk} - \vef\right)
    \\
    &\times
    \int_{\text{BZ}}
    d \bkp
    \delta\left(\ve_{j, \bkp} - \vef\right)
    \int_{\text{MT}} d\br
    \int_{\text{MT}} d\brp
    \\
    &
    \times
    \Psi_{j, \bkp}(\brp)
    \Psi^\ast_{i, \bk}(\brp)
    \Psi^\ast_{j, \bkp}(\br)
    \Psi_{i, \bk}(\br)
    \\
    &
    \times
    \left(\hat{\br} \cdot \hat{\br}^\prime\right)
    \frac{d V_\k(r)}{dr}
    \frac{d V_\k(r^\prime)}{dr^\prime}
    .
    \label{eq:eta_kappa_v2}
  \end{aligned}
\end{equation}

While the explicit use of the wavefunction
expansion from Eq.~\eqref{eq:rmta_wf_expansion}
would lead to four distinct sets of $\{l, m\}$ quantum numbers,
corresponding to each wavefunction, the adoption of the
spherical approximation in Eq.~\eqref{eq:spherical_band}
and the orthonormality of the $\hat{\bk}$- and $\hat{\bk}^\prime$-dependent
spherical harmonics
reduce the number of independent angular momentum indices to two sets:
$\{l, m\}$
and $\{l^\prime, m^\prime\}$.
As a result, Eq.~\eqref{eq:eta_kappa_v2} simplifies to

\begin{equation}
  \begin{aligned}
    \teta_\k &=
    \frac{\Omega^2}{(2 \pi)^6
    N(\vef)
    }
    \sum_{ij}^{N_{\text{states}}}
    \sum_{lm} \sum_{l^\prime m^\prime}
    \\
    &
    \int d k k^2
    |a_{i \k, l}(k)|^2
    \delta\left(\ve_{i, \bk} - \vef\right)
    \\
    &\times
    \int d k^\prime (k^\prime)^2
    |a_{j \k, l^\prime}(k^\prime)|^2
    \delta\left(\ve_{j, \bkp} - \vef\right)
    \\
    &\times
    \left|
    \int_0^{\rmt} d r r^2
    R_{\k, l}(r, \vef)
    \frac{d V_\k(r)}{dr}
    R_{\k, l^\prime}(r, \vef)
    \right|^2
    \\
    &
    \times
    \int d\hat{\br} \int d\hat{\br}^\prime
    Y_{lm}(\hat{\br})
    Y^\ast_{l^\prime m^\prime}(\hat{\br})
    Y^\ast_{lm}(\hat{\br}^\prime)
    Y_{l^\prime m^\prime}(\hat{\br}^\prime)
    \\
    &\times
    \left(\hat{\br} \cdot \hat{\br}^\prime\right)
    .
    \label{eq:eta_kappa_v3}
  \end{aligned}
\end{equation}

The next simplification is achieved by taking
advantage of the separation of the spatial dependence of $\br$ into radial, $r$, and angular, $\hat{\br}$, components.
Applying the identity \cite{Jackson1998}
\begin{equation}
  \begin{aligned}
  \left(\hat{\br} \cdot \hat{\br}^\prime\right)
  =
  \frac{4 \pi}{3}
  \sum_{m''=-1}^{1} Y^\ast_{1 m''}(\hat{\br}^\prime)Y_{1 m''}(\hat{\br}),
  \end{aligned}
\end{equation}
along with the properties of Gaunt coefficients (integrals involving three spherical harmonics \cite{Gaunt1929}),
Eq.~\eqref{eq:eta_kappa_v3}
simplifies to a form involving
only dipole transitions with
$l^\prime = l \pm 1$ and
$m^\prime = m + m''$.
The resulting expression for $\teta_\k$
is given by Eqs.~\eqref{eq:eta_ggp}, \eqref{eq:mll1_def},
and \eqref{eq:nl_def}.

Notably, the partial eDOS, defined in Eq.~\eqref{eq:mll1_def},
can be further simplified by
using the normalization of the radial functions from Appendix~\ref{app:norm_to_dlde}:

\begin{equation}
  \begin{aligned}
    n_{\k, l}&(\rmt; \vef)
    =
    \sum_m n_{\k, lm}(\rmt;
    \vef)
    \\
    &
    =
    \sum_m
    \frac{\Omega \rmt}{(2 \pi)^3}
    \left(
    -
    R_{\k, l}^2
    \frac{\partial L_{\k, l}}
    {\partial \ve}
    \Bigg|_{\substack{r = \rmt\\
    \ve = \vef}}
    \right)
    \\
    &
    \times
    \sum_i
    \int_{\text{BZ}} d \bk
    \delta(\ve_{i, \bk} - \vef)
    \left|a_{i \k, lm}(\bk)\right|^2.
  \end{aligned}
  \label{eq:nl_def_simple}
\end{equation}
In Eq.~\eqref{eq:nl_def_simple}, the integrals over $r$ are replaced
by energy derivatives of the logarithmic derivatives,
${\partial L_{\k, l}}/{\partial \ve}$,
evaluated only on the surface of the MT-sphere.

\section{\label{app:norm_to_dlde}Normalization of the radial functions}

From the radial Schr\"{o}dinger equation
and its energy derivative,
an
integral of the squared radial function can
be converted into the energy-derivative
of the logarithmic derivative of this function
as

\begin{equation}
  \begin{aligned}
  &\int_0^{\rmt}
  dr r^2 R_{\k, l}^2(r, \ve)
  \\
  &=
  \rmt^2
  \left(
  \frac{\partial R_{\k, l}}{\partial \ve} \frac{\partial R_{\k, l}}{\partial r} -
  R_{\k, l} \frac{\partial^2 R_{\k, l}}{\partial r \partial \ve}
  \right)
  \Bigg|_{r = \rmt}
  \\
  &=
  - \rmt
  R_{\k, l}^2
  \frac{\partial L_{\k, l}}
  {\partial \ve}
  \Bigg|_{r = \rmt}.
  \end{aligned}
  \label{eq:useful_identity}
\end{equation}

\section{\label{app:vkappa}Local screened atomic potential}

The coefficients $\tilde{V}(\bG)$ in the plane-wave  expansion of the periodic
local screened
potential $V_{\text{loc}}(\br_0)$
[Fig.~\ref{fig:extract_vkappa}(b)]
are given by Eq.~\eqref{eq:vg},
where the vector $\br_0$ starts at the origin of the
calculated periodic unit cell [see Fig.~\ref{fig:extract_vkappa}(a)].
To obtain the periodic local potential on the atom-centered
$\br$-grid,
with the atom located at $\bm{\tau}_\k$, such that $\br_0 = \bm{\tau}_{\k} + \br$,
one can apply
the inverse Fourier transform

\begin{equation}
  \begin{aligned}
    V_{\k, \text{loc}}(\br) &=
    V_{\text{loc}}(\br_0) =
    V_{\text{loc}}(\bm{\tau}_\k + \br)\\
    &
    =
    \sum_{\bG}
    \left[
    e^{i \bG \cdot \bm{\tau}_\k}
    \tilde{V}(\bG)
    \right]
    e^{i \bG \cdot \br}
    \\
    &
    =
    4 \pi
    \sum_{\bG}
    \left[
    e^{i \bG \cdot \bm{\tau}_\k}
    \tilde{V}(\bG)
    \right]
    \\
    &
    \times
    \sum_{l, m}
    i^l j_l(Gr)
    Y^\ast_{lm}(\hat{\bG})
    Y_{lm}(\hat{\br}).
  \end{aligned}
  \label{eq:vr_ft_pw}
\end{equation}

In the last two lines of Eq.~\eqref{eq:vr_ft_pw}, the
plane wave $e^{i \bG \cdot \br}$
was expanded in terms of
spherical harmonics $Y_{lm}$
and spherical Bessel functions
$j_l$. If only the spherically-symmetric
terms corresponding to
$l = m = 0$ are kept,
Eq.~\eqref{eq:vr_ft_pw}
simplifies
to the
spherically-symmetric atomic potential
given
in Eq.~\eqref{eq:vkappa}.

\section{\label{app:hamann}Semilocal and non-local
parts of Hamann's
pseudopotentials}

For the atom $\k$, the total non-local Vanderbilt-Kleinman-Bylander (NL-VKB)
\cite{Kleinman1982, Bylander1990, Vanderbilt1990}
pseudopotential
corresponding to the angular momentum $l$ is generally
written as

\begin{equation}
  \begin{aligned}
  V^l_{\k, \text{tot}}&(r, r^\prime) =
  V_{\k,\text{loc}}(r)
  +
  V^l_{\k, \text{NL}}(r, r^\prime)
  \\
  &=
  V_{\k,\text{loc}}(r) +
  \sum_{ij \in \{l\}} D_{\k, ij}
  \beta_{\k, i}(r) \beta_{\k, j}(r^\prime),
  \end{aligned}
\end{equation}
where $V^l_{\k, \text{NL}}(r, r^\prime)$
is the fully non-local part that includes
only the beta-projectors $\beta_{\k, i}(r)$ and
$\beta_{\k, j}(r)$ generated for the angular momentum $l$.
These projectors are read directly from the
pseudopotential file along with
the normalization coefficients $D_{\k, ij}$.
The local part
$V_{\k,\text{loc}}(r)$ is extracted from a DFT calculation.
With fully non-local pseudopotentials, the radial Schr\"{o}dinger equation for the radial functions is

\begin{equation}
  \begin{aligned}
    - \frac{d^2}{d r^2}
    \left[r R_{\k, l}(r, \ve)\right]
    &+
    \int d r'
    V^l_{\k, \text{tot}}(r, r^\prime)
    \left[r' R_{\k, l}(r', \ve)\right]
    \\
    &=
    \ve
    \left[r R_{\k, l}(r, \ve)\right].
  \end{aligned}
  \label{eq:rad_schr_nl}
\end{equation}

While
Eq.~\eqref{eq:rad_schr_nl} can be solved numerically,
in some cases it is more straightforward to convert the non-local
pseudopotential to a semilocal form, if such a conversion is possible.
The non-local Hamann's ONCVPSP pseudopotentials
\cite{Hamann1979, Hamann2013},
used in the present work,
allow for this conversion, as the pseudopotential files
include the radial functions $\chi_{\k, l}(r) = r R_{\k, l}(r)$,
evaluated at the same refernce energies
used to construct the beta-projectors.
With these radial functions,
the semilocal components of the potential can be reconstructed
as

\begin{equation}
  \begin{aligned}
    V^l_{\k, \text{SL}}(r) &=
    \frac{1}{\chi_{\k, l}(r)}
    \sum_{ij \in \{l\}} D_{\k, ij} \beta_{\k, i}(r)
    \\
    &\times
    \int d r^\prime \beta_{\k, j}(r^\prime)
    \chi_{\k, l}(r^\prime)
    .
  \end{aligned}
  \label{eq:recover_sl}
\end{equation}

Alternatively, since the Hamann's ONCVPSP code
computes the SL parts internally,
they can be printed directly in the pseudopotential
file and read from there. In our tests,
the SL parts calculated with Eq.~\eqref{eq:recover_sl}
were found to be identical to those printed directly
by the Hamann's ONCVPSP code.

%
%

\bibliography{references}

\end{document}


\title{Rigid muffin-tin approximation in
plane-wave
codes for fast modeling of phonon-mediated superconductors\\
Supplementary Information}%

\author{Danylo~Radevych}%
\email{dradevyc@gmu.edu}
\affiliation{Department of Physics and Astronomy,
George Mason University, Fairfax, VA 22030, USA}
\affiliation{Department of Physics, Applied Physics and Astronomy,
Binghamton University-SUNY,
Binghamton, New York 13902, USA}

\author{Tatsuya~Shishidou}%
\affiliation{Department of Physics,
University of Wisconsin-Milwaukee,
Milwaukee, WI 53201, USA}

\author{Michael~Weinert}%
\affiliation{Department of Physics,
University of Wisconsin-Milwaukee,
Milwaukee, WI 53201, USA}

\author{Elena~R.~Margine}%
\affiliation{Department of Physics, Applied Physics and Astronomy,
Binghamton University-SUNY,
Binghamton, New York 13902, USA}

\author{Aleksey~N.~Kolmogorov}%
\affiliation{Department of Physics, Applied Physics and Astronomy,
Binghamton University-SUNY,
Binghamton, New York 13902, USA}

\author{Igor~I.~Mazin}%
\email{imazin2@gmu.edu}
\affiliation{Department of Physics and Astronomy,
George Mason University, Fairfax, VA 22030, USA}
\affiliation{Quantum Science and Engineering Center, George Mason University,
Fairfax, VA 22030, USA}

\maketitle



\section{Crystallographic parameters}

Crsytallographic parameters of all the structures used in this work
are reported in Table~\ref{tab:crysparam}.


\begin{table}[htbp]
\caption{\label{tab:crysparam}Crystallographic parameters of the considered structures.}
\begin{tabular}{ccccccc}
\hline\hline
\\
Structure &
\multirow{2}{*}{\shortstack{Space\\group}} &
\multirow{2}{*}{\shortstack{$a = b = c$\\(bohr)}} &
\multirow{2}{*}{\shortstack{$\alpha = \beta = \gamma$\\($^\circ$)}} &
Atom &
\multirow{2}{*}{\shortstack{$x$, $y$, $z$\\(crystal units)}} &
\multirow{2}{*}{\shortstack{Wyckoff\\position}}\\
\\
\hline
\\
V (BCC)  & $Im\bar{3}m$ & 5.72 & 90 &
V & 0, 0, 0 & 2a\\
Cu (FCC) & $Fm\bar{3}m$ & 6.83 & 90 &
Cu & 0, 0, 0 & 4a\\
Nb (BCC) & $Im\bar{3}m$ & 6.24 & 90 &
Nb & 0, 0, 0 & 2a\\
Mo (BCC) & $Im\bar{3}m$ & 5.95 & 90 &
Mo & 0, 0, 0 & 2a\\
Rh (FCC) & $Fm\bar{3}m$ & 7.24 & 90 &
Rh & 0, 0, 0 & 4a\\
Pd (FCC) & $Fm\bar{3}m$ & 7.54 & 90 &
Pd & 0, 0, 0 & 4a\\
Ag (FCC) & $Fm\bar{3}m$ & 7.72 & 90 &
Ag & 0, 0, 0 & 4a\\
V$_3$Sn ($A15$)   & $Pm\bar{3}n$ & 9.42 & 90 &
V & $1/4$, $1/2$, 0 & 6d\\
& & & &
Sn & 0, 0, 0 & 2a\\
V$_3$Si ($A15$)   & $Pm\bar{3}n$ & 8.93 & 90 &
V & $1/4$, $1/2$, 0 & 6d\\
& & & &
Si & 0, 0, 0 & 2a\\
Nb$_3$Si ($A15$)  & $Pm\bar{3}n$ & 10.07 & 90 &
Nb & $1/4$, $1/2$, 0 & 6d\\
& & & &
Si & 0, 0, 0 & 2a\\
Nb$_3$Si ($L1_2$) & $Pm\bar{3}m$ & 7.97 & 90 &
Nb & 0, $1/2$, $1/2$ & 3c\\
& & & &
Si & 0, 0, 0 & 1a\\
NbN ($B1$)        & $Fm\bar{3}m$ & 8.29 & 90 &
Nb & $1/2$, $1/2$, $1/2$ & 4b\\
& & & &
N & 0, 0, 0 & 4a\\
\hline\hline
\end{tabular}
\end{table}

\section{Smearing vs. tetrahedron integration methods}

\begin{figure}[htbp]
  \includegraphics[width=1\columnwidth]{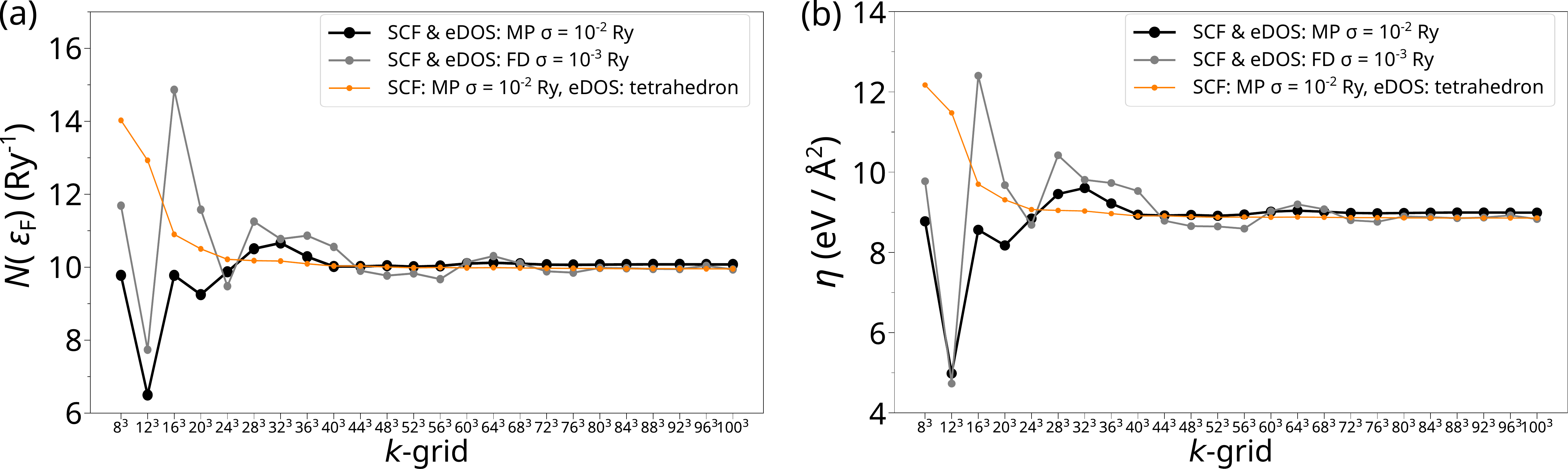}
  \caption{
  Nb BCC:
  (a) eDOS at the Fermi level and (b) McMillan-Hopfield paramater
  convergence with the uniform $k$-grid size depending
  on the integration methods used. ``MP'': Methfessel-Paxton smearing;
  ``FD'': Fermi-Dirac smearing; $\sigma$: corresponding degauss values.
  In all cases, smearing is used for the approximation of the occupation
  function in the SCF part. In the following RMTA calculation,
  eDOS at the Fermi level converges faster when the tetrahedron method
  is used (orange lines)
  instead of the smeared delta-function approximation
  (black and gray lines).
  }
  \label{fig:smearing_nb_bcc}
\end{figure}

Convergence of the total and partial electronic density of states
(eDOS) at the Fermi level
with the uniform $k$-grid size is crucial
to obtain reliable McMillan-Hopfiled parameters.
As can be seen in Figs.~\ref{fig:smearing_nb_bcc}(a),(b),
convergence of the McMillan-Hopfield parameter for Nb BCC
directly follows convergence of eDOS.
When the tetrahedron method of integration is used to
calculate eDOS
through
integrals
with the delta-function in Eqs.~(20),(21) of the main text,
results converge significantly faster
[orange lines in Figs.~\ref{fig:smearing_nb_bcc}(a),(b)]
than
when the same smearing is used for the
approximation of the occupation function in the
self-consistent field (SCF) calculation
and approximation of the delta-function in the eDOS
part
[black and gray lines in Figs.~\ref{fig:smearing_nb_bcc}(a),(b)].
For this reason,
the tetrahedron method of integration
is used by default in all present RMTA calculations.

\section{McMillan-Hopfield parameters compared to previous APW results}
\begin{figure}[htbp]
  \includegraphics[width=0.5\columnwidth]{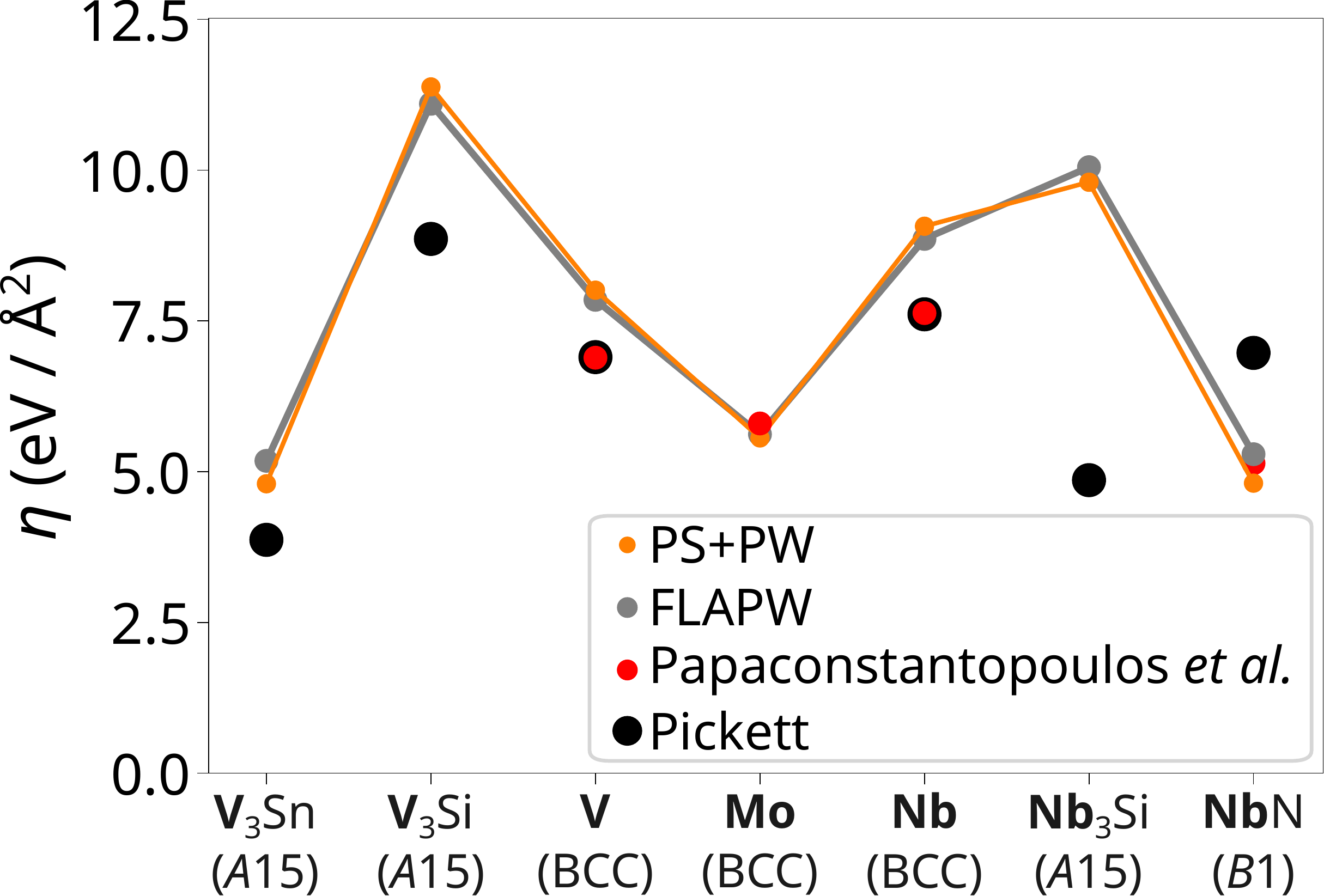}
  \caption{
  Present PS$+$PW and FLAPW McMillan-Hopfield parameters
  for metal atoms--V, Mo, and Nb--in the selected simple metals and compounds compared
  to previous APW results by Papaconstantopoulos \textit{at al.} \cite{Papaconstantopoulos1977} and Pickett \cite{Pickett1982}.
  }
  \label{fig:eta_vs_old_apw}
\end{figure}

In Fig.~\ref{fig:eta_vs_old_apw}, present McMillan-Hopfield parameters
for metal atoms in the selected
simple metals and compounds are compared to results reported by
Papaconstantopoulos \textit{et al.} \cite{Papaconstantopoulos1977} and Pickett \cite{Pickett1982}, who used the APW method and touching MT-spheres.
Our values are close to Papaconstantopoulos \textit{at al.} and Pickett
for simple metals, as well as they are reasonably close to  Pickett's values
for V$_3$Sn, V$_3$Si, and NbN. At the same time, Pickett's value for $A15$-Nb$_3$Si is significantly smaller
than ours.
This could be attributed to various technical details:
differences in crystallographic parameters of the structures,
all-electron potentials, exchange-correlation functionals, and perhaps whether
Nb semicore states were treated as core or valence in the APW method.
Based on the limited technical details provided in the reference \cite{Pickett1982},
it is hard to identify the exact cause(s) of the discrepancy.
All results, however, are correct within the frameworks they are
derived from.

\bibliography{references}